\documentclass[epj,nopacs,final]{svjour}

\usepackage{amssymb}
\usepackage{epsfig}

\begin{document}

\title{Study of QED processes $e^+e^-\to e^+e^-\gamma, e^+e^-\gamma\gamma$ 
with the SND detector at VEPP-2M}

\author{M.N.Achasov  \and  S.E.Baru  \and  A.V.Bozhenok  
\and  A.D.Bukin  \and D.A.Bukin  \and  S.V.Burdin  \and 
T.V.Dimova\thanks{Corresponding author, e-mail: baiert@inp.nsk.su}  \and S.I.Dolinsky  \and 
V.P.Druzhinin  \and M.S.Dubrovin  \and 
I.A.Gaponenko  \and  V.B.Golubev  \and 
V.N.Ivanchenko  \and A.A.Korol  \and  S.V.Koshuba  \and 
A.P.Lysenko  \and I.N.Nesterenko  \and  E.V.Pakhtusova  \and 
S.I.Serednyakov  \and  V.V.Shary  \and 
Yu.M.Shatunov  \and  V.A.Sidorov  \and  Z.K.Silagadze  \and 
Yu.V.Usov }
\institute{Budker Institute of Nuclear Physics, 
Novosibirsk State University, 630090, Novosibirsk, Russia}

\date{Received: date / Revised version: date}

\abstract{
Results of the SND experiment at the VEPP-2M $e^+e^-$ collider
on the QED processes $e^+e^-\rightarrow e^+e^-\gamma$ and
$e^+e^-\rightarrow e^+e^-\gamma\gamma$ with  production at large angles 
are presented.
Energy and angular distributions of the final particles were studied.
No deviations from QED with an accuracy of 3.8\%
for the first process and 10.3\% for the second were found. 
}

\maketitle

\section{Introduction} 
Quantum electrodynamics (QED) describes electromagnetic interactions between
electrons and photons with high accuracy. QED is usually tested in
different types of experiments, for example: 
\begin{itemize}
\item high accuracy ($\leq 10^{-6}$) experiments where high order QED 
corrections at small momentum
transfer are tested, for example,  anomalous magnetic 
moments of leptons,  Lamb shift, etc.;
\item experiments with $e^+ e^-$ colliding beams where QED is tested
at large momentum transfer, for example:
\begin{itemize}
\item[\textbullet] $e^+ e^- \to \gamma \gamma (\gamma \ldots )$,
\item[\textbullet] $e^+ e^- \to e^+ e^- (\gamma,\gamma\gamma  \ldots)$,
\item[\textbullet] $e^+ e^- \to \mu^+ \mu^- (\gamma \ldots )$,
\item[\textbullet] $e^+ e^- \to \tau^+ \tau^- (\gamma \ldots )$.
\end{itemize}
\end{itemize}

This work is devoted to the study of the following QED processes with 
large angles between all particles :
\begin{equation}
e^+ e^- \rightarrow e^+ e^- \gamma , \label{TDeq1}
\end{equation}
\begin{equation}
e^+ e^- \rightarrow e^+ e^- \gamma\gamma . \label{TDeq2}
\end{equation}
This study is important for  several reasons. First,
to check QED as the cross sections and differential distributions
can be precisely calculated and compared with observed ones. 
Second, possible hypothetical leptons,
for example heavy (or excited)
electron \cite{gipel}( the existence of such particle
is ruled out by recent LEP measurements: $m_{e^*}>85-91$ GeV\cite{PDG}), 
can manifest themselves in the invariant mass spectra of
the final particles. Third,
these processes could be a source of background for the vector meson
decays with electrons and photons in the final state. For example,
process (\ref{TDeq2}) is the background in the study of 
decays of $\phi\to\eta e^+e^-,
\eta\to 2\gamma$ and  $\phi\to\eta \gamma$, $\eta\to e^+e^-\gamma$. 
And finally, it is necessary to take into account 
process (\ref{TDeq1}) for  the luminosity measurements  
with accuracy $\sim 1\%$.

The processes  (\ref{TDeq1}) and (\ref{TDeq2}) were studied in different
experiments in different energy regions. Some of these experiments 
are listed in Table \ref{tab1}.
\begin{table}[htb]
\begin{center}
\begin{tabular}{c|c|c}
Experiment & $E_{c.m.}$(GeV) & N.events \\ \hline
\multicolumn{3}{c}{ $e^+ e^- \to e^+ e^- \gamma $} \\ \hline
OLYA\cite{Olya}   & 0.6-1.4 & 1983 \\
ADONE(WAD)\cite{adone}& 1.9-2.9 & 99 \\
CELLO\cite{cello} & 14-46.8 & 934  \\
JADE \cite{jade}  & 34.4 &    3227  \\\hline
\multicolumn{3}{c}{$e^+ e^- \to e^+ e^- \gamma \gamma $} \\ \hline
ND\cite{Ph.Rep}   & 0.6-1.4 & 223 \\
JADE\cite{jade}  & 34.4 &    176 \\\hline
\end{tabular}
\end{center}
\caption{List of some experiments where processes (\ref{TDeq1}) and
(\ref{TDeq2}) were studied.}
\label{tab1}
\end{table}

\section{Detector, experiment}
\begin{figure}[htb]
\epsfig{figure=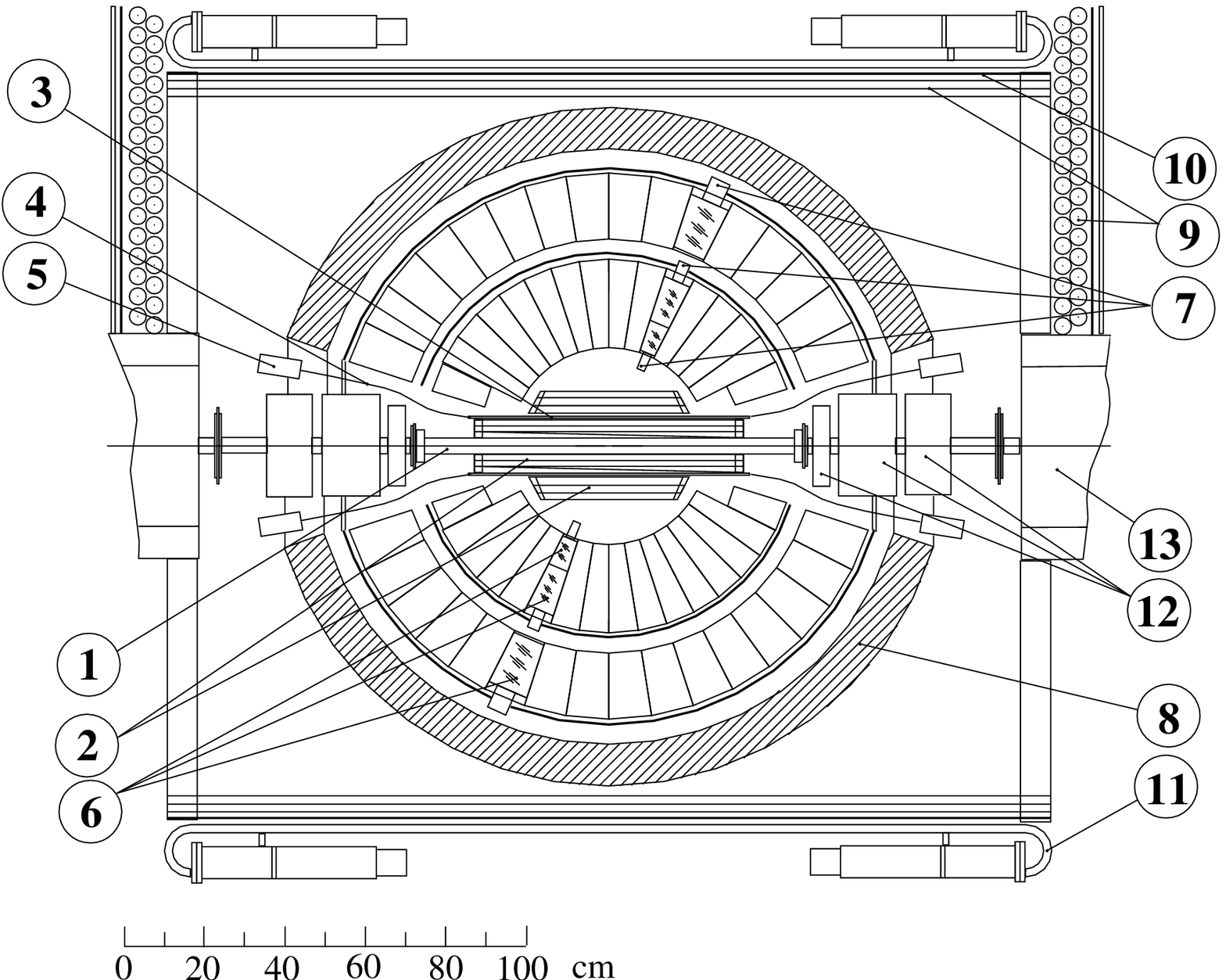,height=5.5cm}
\epsfig{figure=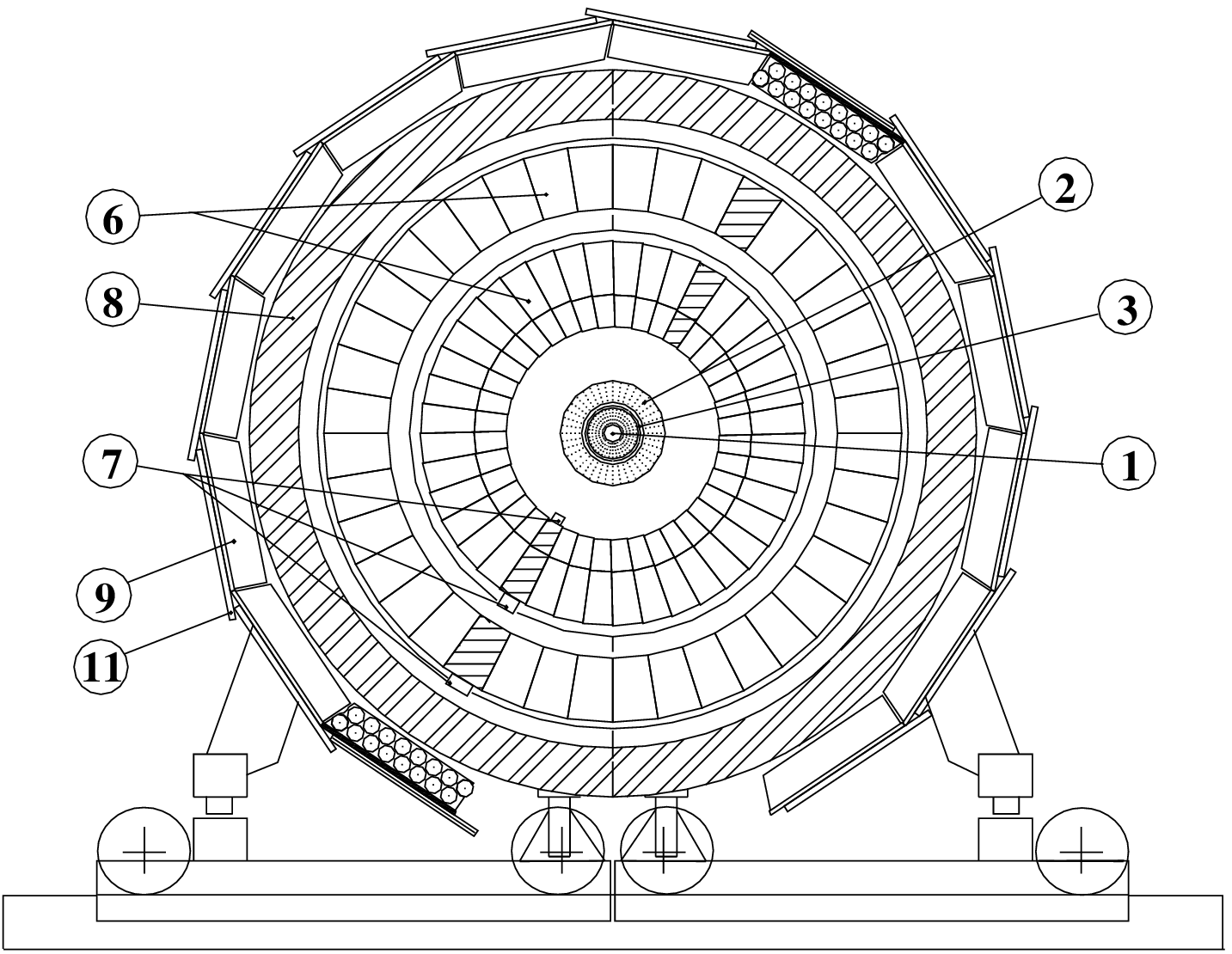,height=5.5cm}
\caption{ SND detector: 1 - beam pipe,
2 -- drift chambers, 3 -- scintillation counter,4 -- light guides, 5 -- PMTs, 
6 -- NaI(Tl) crystals,  7-- vacuum phototriodes, 8 -- iron absorber,
9 -- streamer tubes, 10 -- 1 cm iron plates, 11 -- scintillation counters,
12 and 13 -- collider magnets.}
\label{pic1}
\end{figure}

The experiment \cite{Prep.96,pr97} was carried out with the SND detector
(Fig.\ref{pic1})
at the VEPP-2M collider\cite{VEPP2M}  in the energy region of the
$\phi$-meson resonance  $2E=0.985 - 1.04$ GeV. The SND\cite{SND} detector
is a general purpose
nonmagnetic detector with solid angle coverage $\sim 90\%$ of
$4\pi$. It consists of  a spherical 3 layer
calorimeter based on NaI(Tl) crystals, two drift chambers
and a muon system. 
The list of the SND main parameters is shown in  Table\ref{SND_par}.
The data were recorded in six successive scans 
at 14 different values of the beam energy with the integrated luminosity 
$\Delta L=4.1\mbox{pb}^{-1}$. The accuracy of the luminosity 
determination\cite{pr97} is estimated to be 3\%.

\begin{table}
\caption{\label{SND_par} \bf List of SND parameters}
\begin{tabular}{|ll|}  \hline
{\bf Calorimeter:}  & \\
Total number of NaI(Tl) counters & 1632 \\
Angular size of the counter & $\Delta \varphi = \Delta\vartheta = 9^\circ$ \\
Readout & vacuum  \\
        & phototriodes \\
Noise per one counter & $\sim $ 0.3 MeV \\
Energy deposition from $\gamma$'s $\Delta E_\gamma $ &
$ (0.91 \pm 0.02) E_0 $ \\
($E_\gamma = 50 -700$ MeV) & \\

Energy resolution for $\gamma$'s\cite{calibr} $\sigma E_\gamma / E_\gamma$ &
$4.2\%/(E(GeV))^{1/4}$ \\

Angular resolution for $\gamma$'s  &
$\delta\varphi = \delta\vartheta = 1.5^\circ$ \\
& $(E_\gamma =300\mbox{\ MeV})$\\

Minimal spatial angle for two photons  &
$\Delta\varphi \sim \Delta\vartheta\sim 18^\circ$ \\
separation & \\
\hline

{\bf Drift chambers} : & \\

Spatial resolution for tracks &
 $\sigma _\varphi =0.3^\circ$,   \\
(P=300 MeV/c) &  $\sigma _\vartheta =2.5^\circ$ \\

Minimal azimuth angle for charged   & 
$\Delta\varphi \sim 18^\circ$ \\
particles separation &\\
Amount of material before the chamber & 0.27  $g/cm^2$ \\
Probability of $\gamma$-conversion before  & 0.57\% \\
the chamber & \\
\hline
\end{tabular}
\end{table}

\section{Simulation}

Monte Carlo simulation was used for comparison
of the experimental results with theoretical predictions. 
Full simulation of the detector  was made on the base of
UNIMOD2 program \cite{UNIMOD2}. 
The process (\ref{TDeq1})
was simulated according to formulae of the $\alpha^3$ order from 
Ref.\cite{Baier}. The details of the implementation of these formulae into
event generator program are described in Ref.\cite{TDUM}.

For the process (\ref{TDeq2})
formulae of the $\alpha^4$ order of differential
cross section, calculated with the method of helicity amplitudes \cite{TDEEGG}
were used. These formulae are valid  when all angles
between final particles are large. So the simulation was performed under
a condition that all angles are larger than $15^\circ$. 

The radiative correction for process (\ref{TDeq1})
was calculated using formulae from Ref.\cite{Kuraev}.
The corrected cross section can be written as:
$\sigma_{th} = \sigma_{B}(1+\delta)$, where
$\sigma_{B}$ is an $\alpha^3$ Born cross section and $\delta$ -
calculated radiative correction.
The radiation of virtual and  soft photons as well as
hard  photon emission close  to the direction of motion
of one of the initial or final charged particles were taken
into account. These formulae were integrated over phase
space as close as possible to the experimental acceptance.
The decrease in the registration efficiency due to
lost radiative photon was taken into account in calculations
of contribution from hard photon radiation. 
As a  result  $\delta  = - (10 \pm 3)\%$  was obtained.
The error originates from two main
sources: the formula for differential cross section of virtual
and soft photon radiation corrections is incomplete ( $\sim 3\%$),
estimation of the efficiency dependence due to the loss of radiative
photon ($\sim 1\%$).

\section{Data Analysis}

At the first stage of data analysis the following selection criteria,
common for both processes, were applied:
\begin{itemize}
\item number of charged particles $ N_{cp}=2$ 
\item number of photons $ 1\leq N_\gamma \leq 3 $
\item both tracks originate from the interaction region: 
distance between tracks and beam axis
in $R - \phi$ plane $R_{1,2} < 0.5$cm, Z coordinate of the 
closest to the beam axis point on the track
$|Z_{1,2}| < 10$ cm
\item polar angles of all particles $ 36^o < \theta < 144^o$
\item acollinearity angle of charged particles in the plane
transverse to the beam axis $|\Delta \phi _{ee}| = 
|180^o - |\phi_1-\phi_2|| > 5^o$
\item normalized total energy deposition $E_{tot}/2E_0 > 0.8$
\item normalized total momentum$ P_{tot}/E_{tot} < 0.15 $
\item minimal energy of charged particle $E_{e~min} > 10$ MeV
\item minimal energy of photon $E_{\gamma~min} > 20$ MeV
\item no hits in muon system
\end{itemize}

Nearly 90000 events passed these cuts for use in  further analysis.

\subsection{Process $e^+e^-\rightarrow e^+ e^-\gamma$}
For the selection of events from process (\ref{TDeq1}) 
a kinematic fit imposing 4-momentum conservation was applied.
The   parameter $\chi^2$, describing 
the degree of energy-momentum balance in the event, was calculated.
For the selection of events from the process $e^+e^-\rightarrow e^+e^-\gamma$
an additional cut was imposed:
\begin{itemize}
\item[\textbullet]  $\chi^2 < 15 $
\end{itemize}

The number of thus selected events in experiment and
simulation  of process (\ref{TDeq1}) as well as for some background processes
are shown in the Table \ref{tab2}.

\begin{table}[htb]
\begin{center}
\begin{tabular}{|c|c|c|c|}
\hline
&Number & Detection  & Visible \\
Process& of   &efficiency& cross \\
&events & (\%) & section \\ \hline
$e^+ e^- \gamma$(Exp)&73692& & 17.9$\pm$0.1 nb \\
$e^+ e^- \gamma$(MC) &6081&59.8$\pm$1.0 & 19.7$\pm$ 0.3 nb \\
$\omega \pi^0$ (MC)&   1 & 0.0033 & $\sim$ 0.0003 nb \\
$\pi^+ \pi^- \pi^0$(MC)&556& 0.19 & $\sim$ 0.02 nb \\ 
$\pi^+ \pi^- \gamma$(MC)& 8 & 0.08 & $\sim$ 0.05 nb \\ \hline
\end{tabular}
\end{center}
\caption{Number of events which passed the selection criteria for
$e^+e^-\to e^+e^-\gamma$ and background processes}
\label{tab2}
\end{table}

\begin{figure}[htb]
\epsfxsize90mm
\epsfbox{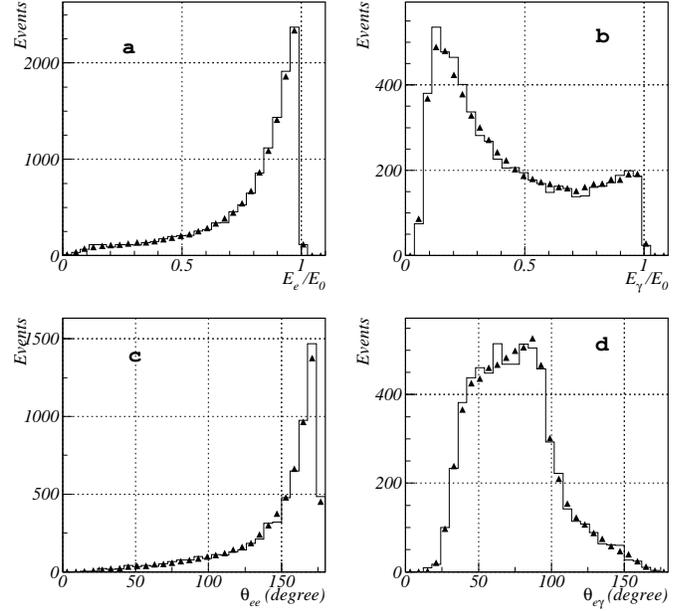}
\caption{Energy and angular spectra for the process $ e^+e^-\to
e^+e^-\gamma$: a) energy spectrum of charged particles
b) energy spectrum of photons c) angle  between
charged particles d) minimal angle between charged particle and
photon;
$\blacktriangle$ -- experimental points, histogram -- simulation. }
\label{pic2}
\end{figure}

\begin{figure}[htb]
\epsfxsize90mm
\epsfbox{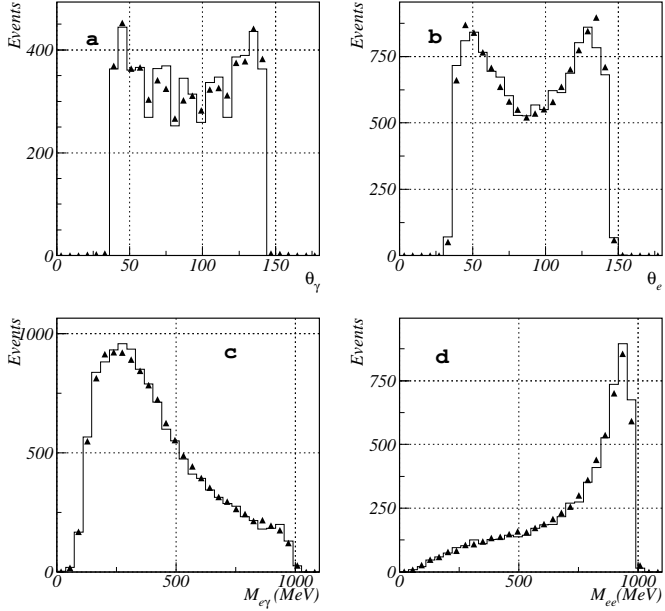}
\caption{Angular and invariant mass  spectra for the process $ e^+e^-\to
e^+e^-\gamma$: a) polar angle of photons b)polar angle of charge
particles c) invariant mass of pair consisting of charged particle plus photon
d) invariant mass of charged particles;
$\blacktriangle$ -- experimental points, histogram -- simulation. }
\label{pic3}
\end{figure}

The corresponding energy, angular and invariant mass distributions after
kinematic fit are shown
in Fig.\ref{pic2},\ref{pic3}. The statistical errors in these figures 
are comparable with the
marker size. The peaks  in  Fig.\ref{pic2}a,b,c originate  from 
quasi-elastic events
of process (\ref{TDeq1}) with  radiation of a soft photon
with energy $E_{\gamma}/E_0\ll 1$. There is good
agreement between experimental data and MC simulation. There are no traces
of heavy lepton in the invariant mass spectrum in Fig.\ref{pic3}c.
Some minor differences
in the spectra (Fig.\ref{pic2}d, \ref{pic3}a) could be 
attributed to imprecise simulation of angular differential nonlinearity
for photons caused by granularity of the calorimeter.

The estimated detection efficiency for the described selection criteria is equal to
$59.8 \pm 1.0\%$ (error is statistical).
It was defined with respect to simulation under following the
conditions: polar angle of final  particles $36^\circ < \theta < 144^\circ$,
azimuth
acollinearity angle $ \Delta \phi_{ee} >5^\circ$,
spatial angle between final particles is 
$\theta_{ee,e\gamma}>20^\circ$
minimal energies for charged particles
and photons are equal to 10 and 20 MeV respectively. 
The systematic error on the measured cross section 
is determined by normalization uncertainty  (3\%), limited MC statistics
(1.7\%) and uncertainties in the selection efficiency
(1.5\%). In total it is equal to 3.8\%. 

\begin{figure}[htb]
\epsfxsize90mm
\epsfbox{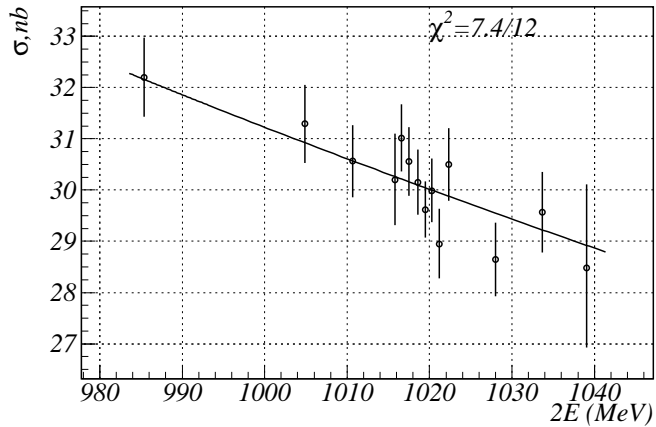}
\caption{ Cross section energy dependence for the process $ e^+e^-\to
e^+e^-\gamma$. Points - experiment, line - fit with  formula
(\ref{TDeq3}). }
\label{pic4}
\end{figure}

 The energy dependence of the cross section of process
(\ref{TDeq1}) is shown in Fig.\ref{pic4}. The measurements were fitted
using the following function:
\begin{equation}
\sigma (E)= \sigma_0 (E)\cdot (E_0^2/E^2) + W \cdot\sigma_{\phi}(E),
\label{TDeq3}
\end{equation}
where the first term has the energy dependence typical of QED processes
and the second corresponds to a contribution from $\phi$-meson decays 
with cross section $\sigma_\phi$. The
fitting parameters are
$\sigma_0$ --- the cross section at the energy
$E_0=1020 $MeV and $W$ determines resonance background contribution.
The main part of this  background for process \ref{TDeq1} comes from
$\phi\to\pi^+\pi^-\pi^0$ decay.

Fitting gives  no peak from $\phi$-meson decays
(fig.\ref{pic4}). 
The fitted  experimental cross section is
$\sigma_0 = 30.01\pm 0.12 \pm 1.2 $ nb  and the expected QED cross section
with  radiative corrections is
$\sigma_{th} = 29.7\pm 0.3 \pm 1.0$ nb. The observed difference 
($\sim$ 1\%) is within  the systematic error.

\subsection{Process $e^+e^-\rightarrow e^+ e^-\gamma\gamma$}
For the selection  of events from  the process 
$e^+e^-\rightarrow e^+e^-\gamma\gamma$,
the following additional cuts were imposed:
\begin{itemize}
\item[\textbullet] number of photons $ 2\leq N_\gamma \leq 3 $,
\item[\textbullet]  $\chi^2 < 15 $,
\item[\textbullet] to suppress the
contribution from $e^+e^-\to \pi^+\pi^-\pi^0$
region $ 110 <M_{\gamma\gamma} <170 $MeV was excluded,
\item[\textbullet] minimal energy of photons $E_{\gamma~min} = 50$ MeV.
\end{itemize}
Here $\chi^2$ - is the kinematic fit  parameter obtained
under the assumption that events come from  process (\ref{TDeq2}).
The number of events which passed these selection criteria in the experiment and
Monte Carlo simulation of process (\ref{TDeq2}) and background processes
are shown in  Table \ref{tab3}.  

\begin{table}[htb]
\begin{tabular}{|c|c|c|c|}
\hline
   & Number  & Detection & Visible \\
Process & of &efficiency & cross \\
     & events   &   (\%)   & section(nb) \\ \hline
$e^+ e^- \gamma\gamma$ (Exp.) & 698 & & 0.153 $\pm$ 0.013  \\
$e^+ e^- \gamma\gamma$ (MC)   & 647 & 33.6$\pm$ 1.5 & 0.151$\pm$ 0.006  \\
$ \omega \pi^0$ (MC) &      3   & 0.01 & $\sim$ 0.001   \\
$ \pi^+ \pi^- \pi^0$ (MC)& 16  & 0.006 & $\sim$ 0.0006  \\
$ \pi^+ \pi^- \gamma$ (MC)& 1  & 0.001 &$\sim$ 0.001  \\ \hline
\end{tabular}
\caption{Number of events which passed the selection criteria for
$e^+ e^- \to e^+ e^-\gamma\gamma$ and background processes.}
\label{tab3}
\end{table}

\begin{figure}[htb]
\epsfxsize90mm
\epsfbox{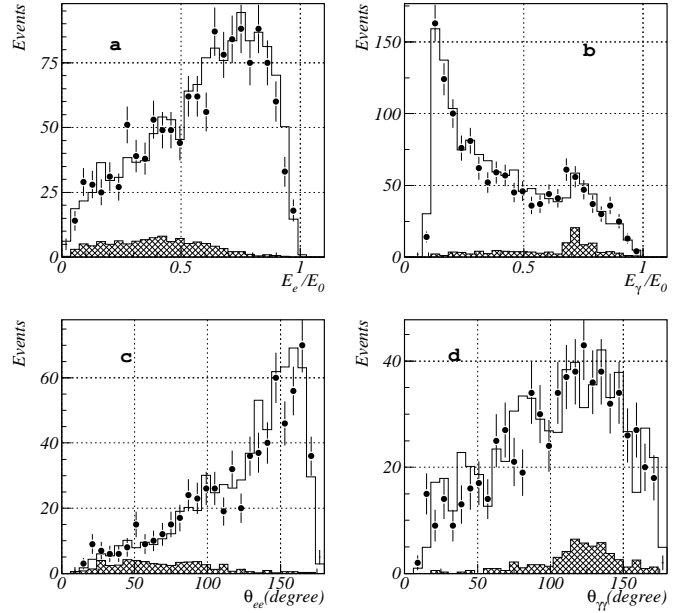}
\caption{  Energy and angular spectra for the process $ e^+e^-\to
e^+e^-\gamma\gamma$: a) energy spectrum of charged particles
b) energy spectrum of photons c) angle  between
charged particles d) angle between photons;
$\bullet$ -- experimental points, filled histogram - simulation of background
from Dalitz decays $\phi\to\eta e^+e^-, \eta\to \gamma\gamma$ and
$\phi\to\eta\gamma, \eta\to e^+e^-\gamma$, histogram -- sum of simulations
of QED process and background.}
\label{pic5}
\end{figure}

\begin{figure}[htb]
\epsfxsize90mm
\epsfbox{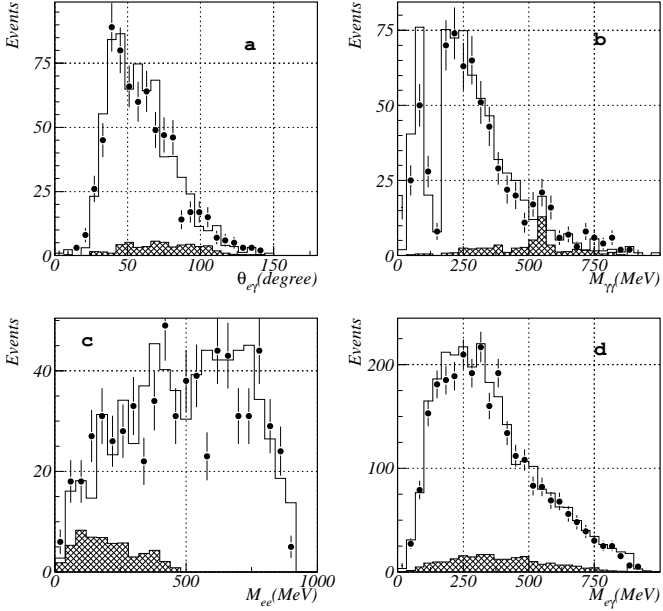}
\caption{ Angular and invariant mass  spectra for the process $ e^+e^-\to
e^+e^-\gamma\gamma$: a) minimal angle between charged particles
and photons b) invariant mass of two photons
c) invariant mass of charged particles,
d) invariant mass of pair charged particle and photon;
$\bullet$ -- experimental point, filled histogram - simulation of background
from Dalitz decays $\phi\to\eta e^+e^, \eta\to \gamma\gamma$ and
$\phi\to\eta\gamma, \eta\to e^+e^-\gamma$, histogram -- sum of simulations
of QED process and background.}
\label{pic6}
\end{figure}

Energy, angular and invariant mass distributions after kinematic fit are
shown in Fig.\ref{pic5},\ref{pic6}. Similar to process (\ref{TDeq1})
the peaks are seen from quasi-elastic
scattering with emission of soft photons (Fig.\ref{pic5}a,b,c). The peak
in photon energy spectra (Fig.\ref{pic5}b) near $E_\gamma/E_0=0.7$
corresponds to the recoil photon energy in
radiative decays: $\phi\to\eta\gamma$, $\eta\to e^+e^-\gamma,
\pi^+\pi^-\gamma$.
Some enhancement in the 
two photon invariant mass spectrum  (Fig.\ref{pic6}b) near  the 
$\eta$-mass appears from the 
decay $\phi\to\eta e^+e^-, \eta\to\gamma\gamma$. There are also no visible 
traces of
heavy lepton production in the $M_{e\gamma}$ spectrum (Fig.\ref{pic6}d).

The detection  efficiency  was determined from simulation in 
nearly the  same conditions
as for process (\ref{TDeq1}) :
 polar angle of final  particles $36^\circ < \theta < 144^\circ$, azimuth
acollinearity angle  $ \Delta \phi_{ee}>5^\circ$,
spatial angle between final particles
$\theta_{ee,e\gamma,\gamma\gamma}>20^\circ$,
minimal energies for charged particles
and photons are equal to 10 and 50 MeV respectively,
The value of detection efficiency was found to be 33.6$\pm$ 1.5\%.

\begin{figure}
\epsfxsize90mm
\epsfbox{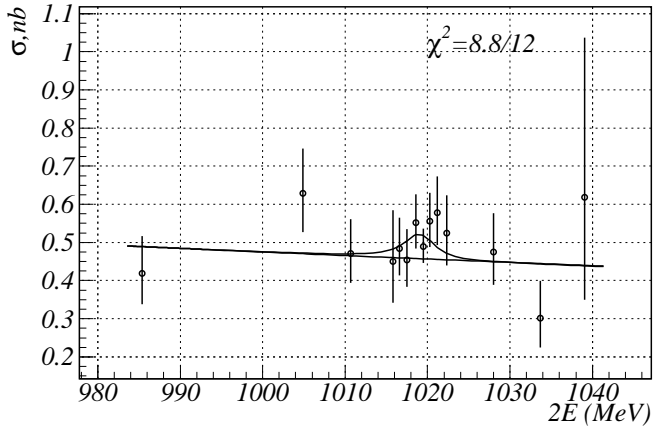}
\caption{ Cross section energy dependence for the process $ e^+e^-\to
e^+e^-\gamma\gamma$. Points - experiment, line - fit with formula
(\ref{TDeq3}). }
\label{pic7}
\end{figure}

The fitting of the energy dependence of the cross section of
process (\ref{TDeq2}) was done using formula (\ref{TDeq3}).
The result is shown in Fig.\ref{pic7}. The contribution from $\phi$ decays
is seen  as  a peak at  the $\phi$ mass. The significance of the peak
is $\sim$ 1.5 of standard deviation. The processes
$\phi\to\eta e^+e^-, \eta\to \gamma\gamma$ and
$\phi\to\eta\gamma, \eta\to e^+e^-\gamma$, mentioned above,
constitute the main contribution to the peak.
The fitted value of experimental cross section
$\sigma_0 = 0.457\pm 0.039 \pm 0.026 $ nb
was found  in good agreement with the
calculated QED cross section  $\sigma_{MC} = 0.458\pm 0.010$ nb.
The systematic error included into $\sigma_0$ 
is determined by normalization uncertainty  (3\%), limited MC statistics
(4.5\%) and uncertainties on the selection efficiency
(2.\%). In total it is equal to 5.8\%. 

\section{Conclusions}
In the experiment with the SND detector at the VEPP-2M collider 
the $e^+e^-\to e^+e^-\gamma$ and  $e^+e^-\to e^+e^-\gamma\gamma$
QED processes
with particles produced at large angles were studied. 
A total of 73692 events of the process $e^+e^-\to
e^+e^-\gamma$ was observed. 
For the process $e^+e^-\to e^+e^-\gamma\gamma$
698 events were observed where 649 events are from the QED process
(\ref{TDeq2}).
Number of events observed in different energy points
for both processes are shown in tables \ref{tab4},\ref{tab5}.
The cross sections and differential distributions of produced
particles were compared with
MC simulation. No significant deviations 
from  QED were found  within limits of  measurement errors, which
are equal to 3.8\%  and 10.3\% for processes (\ref{TDeq1}) and (\ref{TDeq2})
respectively.

\begin{table}[htb]
\begin{center}
\begin{tabular}{|c|c|c|}
\hline
$E_{c.m.}$(MeV) & N.events & Experimental  \\ 
                &          & cr.section(nb)\\ \hline
 985.4& 3827  & 19.26 \\
1004.9& 3545  & 18.71 \\
1010.7& 4616  & 18.27 \\
1015.8& 2150  & 18.06 \\
1016.6& 5811  & 18.54 \\
1017.5& 5506  & 18.27 \\
1018.6& 7492  & 18.03 \\
1019.5& 19049 & 17.71 \\
1020.3& 7395  & 17.93 \\
1021.2& 4550  & 17.31 \\
1022.3& 4138  & 18.24 \\
1028.0& 3436  & 17.13 \\
1033.7& 2792  & 17.67 \\
1039.1&  410  & 17.03 \\ \hline
\end{tabular}
\end{center}
\caption{ Number of events and experimental cross section
for $e^+e^-\to e^+e^-\gamma$.}
\label{tab4}
\end{table}

\begin{table}[htb]
\begin{center}
\begin{tabular}{|c|c|c|}
\hline
$E_{c.m.}$(MeV) & N.events & Experimental  \\ 
                &          & cr.section(pb)\\ \hline
 985.4& 28  & 141. \\
1004.9& 40  & 210. \\
1010.7& 40  & 156. \\
1015.8& 17  & 144. \\
1016.6& 48  & 153. \\
1017.5& 42  & 139. \\
1018.6& 68  & 163. \\
1019.5& 156 & 145. \\
1020.3& 70  & 169. \\
1021.2& 48  & 182. \\
1022.3& 39  & 170. \\
1028.0& 32  & 159. \\
1033.7& 16  & 101. \\
1039.1&  5  & 208. \\ \hline
\end{tabular}
\end{center}
\caption{ Number of events and experimental  cross section
for $e^+e^-\to e^+e^-\gamma\gamma$ with subtracted $\phi-meson$ 
background.}
\label{tab5}
\end{table}

\begin{acknowledgement}
{\it Acknowledgements.} This work was supported in part  by Russian Foundation of 
Basic Researches (grant No.96-15-96327); and STP ``Integration'' (Grant No 274).
\end{acknowledgement}

\end{document}